\documentclass[runningheads]{llncs}

\usepackage{graphicx}
\usepackage{subcaption}
\usepackage{todonotes}
\usepackage{booktabs}       
\usepackage{cite}
\usepackage{float}
\usepackage{makecell}
\usepackage{lipsum}

\usepackage{amsmath}

\begin{document}
%
\title{nnUNet\_RASPP for Retinal OCT Fluid Detection, Segmentation  and Generalisation over Variations of Data Sources}
%
\titlerunning{nnUNet\_RASPP for Retinal OCT Fluid Segmentation}

%
\author{Nchongmaje Ndipenoch, Alina Miron, Zidong Wang and Yongmin Li}
\institute{Department of Computer Science, Brunel University London, Uxbridge, UB8 3PH, United Kingdom}


\maketitle              

\begin{abstract}
Retinal Optical Coherence Tomography (OCT), a noninvasive cross-sectional scan of the eye with qualitative 3D visualization of the retinal anatomy is use to study the retinal structure and the presence of pathogens. 
The advent of the retinal OCT has transformed ophthalmology and it is currently paramount for the diagnosis, monitoring and treatment of many eye pathogens including Macular Edema which impairs vision severely or Glaucoma that can cause irreversible blindness. 
However the quality of retinal OCT images varies among device manufacturers. Deep Learning methods have had their success in the medical image segmentation community but it is still not clear if the level of success can be generalised across OCT images collected from different device vendors. In this work we propose two variants of the nnUNet \cite{isensee:NPG2021}. The standard nnUNet and an enhanced vision  call nnUnet\_RASPP (nnU-Net with residual and Atrous Spatial Pyramid Pooling) both of which are robust and generalise with consistent high performance across images from multiple device vendors. The algorithm was validated on the MICCAI 2017 RETOUCH challenge dataset \cite{bogunovic:IEEE2019} acquired from 3 device vendors across 3 medical centers from patients suffering from 2 retinal disease types. Experimental results show that our algorithms outperform the current state-of-the-arts algorithms by a clear margin for segmentation obtaining a mean Dice Score (DS) of 82.3\% for the 3 retinal fluids scoring 84.0\%, 80.0\%, 83.0\%  for  Intraretinal Fluid (IRF), Subretinal Fluid (SRF), and Pigment Epithelium Detachments (PED) respectively on the testing dataset. Also we obtained a perfect Area Under the Curve (AUC) score of 100\% for the detection of the presence of fluid for all 3 fluid classes on the testing dataset.

\end{abstract}

\section{Introduction}
\label{section:introduction}
Macular Edema is an eye condition that occurs when there is a leakage of blood vessels into part of the retinal called Macular (central of the eye at the back where the vision is sharpest) and hence impairing the vision severely. There are many eye diseases that can cause this including, Age-related Macular Degeneration (AMD) and Diabetic Macular Edema (DME). Recent study \cite{li:EURETINA2017} indicates that there's a rise of retinal diseases in Europe with more than 34 and 4 million people affected with AMD and DME respectively in the continent.
AMD, is mostly common among older people (50-years and above). The early stage of AMD is asymptomatic and slows to progress to the late stage which is more severe and less common.
DME, is the thickening of the retinal caused by the accumulation of intraretinal fluid in the macular and it's mostly common among diabetic patients.
Currently there is no cure for these diseases and anti-vascular endothelial growth factor (Anti-VEGF) therapy is the main treatment. This requires constant administering of injections which are expensive and hence a socio-economic burden to most patients and the healthcare system.
Therefore early diagnostic and active monitoring the progress of these diseases is vital because the doctors can give some behavioral advice like change of diet or doing regular exercise which will help slow down the progress or in some cases prevents the diseases from getting into a later stage. As of today this is mostly done manually which is laborious, time intensive and prone to error. Therefore an automatic and reliable tool is very crucial in this process and to further exploit the qualitative features of the retinal OCT modality efficiently.
Also, the presence of eye motion artifacts in OCT lowers the signal-to-noise ratio (SNR) due to speckle noise. To circumvent this problem device manufacturers have to find a balance between achieving high SNR, image resolution and the scanning time. Hence the quality of the images varies among device vendors and hence the need to develop an automate tool with high performance that can generalise across images from all the device vendors. 

To address the above issues, in this work we propose the nnUNet \cite{isensee:NPG2021} and an enhanced version call nnUnet\_RASPP.
Our main contribution is enhancing the nnUNet by integrating residual blocks and an Atrous Spatial Pyramid Pooling (ASPP) block to the network's architecture.

The rest of the paper is organized as follows. A brief review of the previous studies is provided in Section \ref{section:background}. Section \ref{section:methods} presents the proposed methods. The experiment with results and visualisation are presented in Section \ref{section:experiments}, and  finally, the conclusion with our contributions is described in Sections \ref{section:conclusions}. 

\section{Background}
\label{section:background}

OCT was first developed in the early 1990s but only became commercially available in 2006 and rapidly became popular due to its high image quality resolution. The segmentation of retinal OCT images have been around for many years from graph-cut\cite{Salazar:jbhi2014, Salazar:icarcv2010}, Markov Random Fields \cite{Salazar:his2012, Wang:jbhi2017}, level set \cite{Dodo:access2019,Dodo:bioimaging2019}, to the recent Deep Learning methods that will be briefly reviewed as below. 

Unet, a Deep Learning approach for medical image segmentation is introduced in \cite{Ronneberger:MICCAI2015} by Olaf Ronneberger et al. Like the name: Unet suggest, the architecture has a U-shape and consists of an encoder, bottleneck and a decoder block. It's an end to end framework in which the encoder is use to extract features from the input images/maps, and the decoder is used for pixels localisation. At the end of the decoder path is a classification layer to classify each of the pixels to belong to each of the segmented class. Also, between the encoder and the decoder paths is a bottleneck to ensure the smooth transition from the encoder to the decoder. The encoder, decoder and bottleneck are made up of a series of convolutional layers arranged in a special order.

The Deep-ResUNet++ is presented in \cite{ndipenoch:CISP-BMEI2022} for simultaneous segmentation of layers and fluids in retinal OCT images. The approached incorporated residual connections, ASPP blocks and Squeeze and Exciting blocks into the traditional 2D Unet \cite{Ronneberger:MICCAI2015} architecture to simultaneously segment 3 retinal layers, 3 fluids and 2 background classes from 1136 B-Scans from 24 patients suffering from wet AMD. The algorithm is validated on the  Annotated Retinal OCT Images (AROI) \cite{Melinscak:Automatika2021} which is publicly available.

A clinical application for diagnosis and referral of retinal diseases is proposed in \cite{Fauw:NatureMedicine2018} in which 14,884 OCT B-Scans collected from 7,621 patients are trained on a framework consisting of two main parts : The segmentation model (3D Unet \cite{Cicek:MICCAI2016}) and the classification model. 

An approach to segment fluids from retinal OCT using Graph-Theory (GT) and Fully Convolutional Networks (FCN) with curvature loss is presented in \cite{xing:IEEETMI2022}. The GT is used to delineate the retinal layers, the FCN for segmentation and the loss function further uses curvature regularization term to smooth boundary and eliminate unnecessary holes inside the predicted fluid. The algorithm was validated on the RETOUCH dataset \cite{bogunovic:IEEE2019} consisting of 3 fluid types. 

A combination of Convolutional Neural Networks (CNN) and Graph Search (GS) method is presented in \cite{Fang:BOE2017}. The framework aims to validate nine layers boundaries from 60 retinal OCT volumes (2915 B-scans, from 20 human eyes) obtained from patients suffering from dry AMD. CNN is used for the extraction of the layer boundaries features while the GS is used for the pixels classification.

In \cite{pekala:CBM2019} another Deep Learning approach for retinal OCT segmentation combining a FCN for segmentation with Gaussian Processes for post processing is proposed. The method is validated on the University of Miami dataset \cite{tian:JOB2016} which consists of 50 volumes from 10 patients suffering from diabetic retinopathy. Their approach is divided into two main steps which are the pixel classification using the FCN and the post processing  using Gaussian Processes.

Another CNN-based approach for the simultaneous segmentation of layers and fluid is presented in \cite{Roy:BOE2017}. They presented a 2D Unet like architecture with a reduced depth for the segmentation of 10 classes consisting of 8 layers, 1 background and 1 fluid from 10 patients suffering from Diabetic Macular Edema (DME). The Duke DME dataset \cite{Chiu:BOE2015} is used to validate the algorithm.

A 3-part CNN-based and Random Forest (RF) framework was developed by \cite{Lu:arXiv:2017} to segment and detect fluids in OCT images. The first part of the framework is used for pre-processing of the images, the second part consists of a 2D Unet architecture for the extraction of features and a RF classifier is used at the third part to classify the pixels. The framework is validated on the MICCAI 2017 RETOUCH challenge dataset \cite{bogunovic:IEEE2019}.

A combination of CNN and graph-shortest path (GSP) method is presented in \cite{rashno:MICCAI2017} for the segmentation and detection of fluid in retinal OCT images. The algorithm  is validated on the MICCAI 2017 RETOUCH challenge dataset \cite{bogunovic:IEEE2019}. In this method the CNN is used for the segmentation of region of interest (ROI) and the GSP is further used for the segmentation of the layers and fluid from the ROI.

A standard double-Unet architecture for the detection and segmentation of fluids in retinal OCT images is proposed in \cite{kang:MICCAI2017}. The method uses 2 Unet architectures connected in series and is validated on the MICCAI 2017 RETOUCH challenge dataset. The output of the first Unet serves as an input to the second Unet.

nnUNet, a self-configuring framework is introduced in \cite{isensee:NPG2021}. The framework aims to eliminate the problem of manual parameters setting "trying an error" by using the dataset's demographic features to determine and automatically set some of the model's key parameters like the batch size. The framework uses the standard Unet \cite{Ronneberger:MICCAI2015} and is evaluated on 11 biomedical image segmentation challenges consisting of 23 datasets for 53 segmentation tasks.

An extended version of the nnUNet \cite{isensee:NPG2021} is presented by McConnell et al in \cite{mcconnell2:CBMS2022} by integrating residual, dense, and inception blocks into the network for the segmentation of medical imaging on multiple datasets. The algorithm is evaluated on eight datasets consisting of 20 target anatomical structures.

ScSE nnU-Net, another extended version of the nnUNet \cite{isensee:NPG2021} is presented in \cite{xie:MICCAI2020} for the segmentation of head and neck cancer tumors. It extends the original nnUnet by incorporating spatial channels with squeeze and excitation blocks into the network's architecture. The algorithm uses nnUNet to extract features from the input images/maps and then the squeeze and excitation blocks to further suppress the weaker pixels. The method was validated on the HECKTOR 2020 training dataset consisting of 201 cases and a test set of 53 cases.

In the medical image segmentation community the Unet is the most common and widely used architecture but most of the parameters are setup manually "try an error". Therefore we aim to improve the performance of the Unet by leveraging on nnUNet, a self-parameterise pipeline for medical image segmentation and adapt it to solve the problem of data source variance as explain in the next part of this paper.


\section{Methods}
\label{section:methods}
nnUnet\_RASPP is inspired and adapted from nnUNet \cite{isensee:NPG2021} developed by Isensee et al. It is a self-configuring and automatic pipeline for medical image segmentation which helps to mitigate the problem of manual parameters setting "try an error".
We enhanced the nnUNet by incorporating residual and ASPP blocks into the network's architecture. 
In this section we will give a brief summary of the standard Unet \ref{subsection:UNet}, nnUNet \ref{subsection:nnUNet}, residual connections \ref{subsection:residual_conne}, ASSP \ref{subsection:aspp} block and finally explain how we integrate these components to build nnUnet\_RASPP \ref{subsection:nnUnet_RASPP}.


\subsection{Unet}
\label{subsection:UNet}

\begin{figure}[H]
\centerline{\includegraphics[width=11cm]{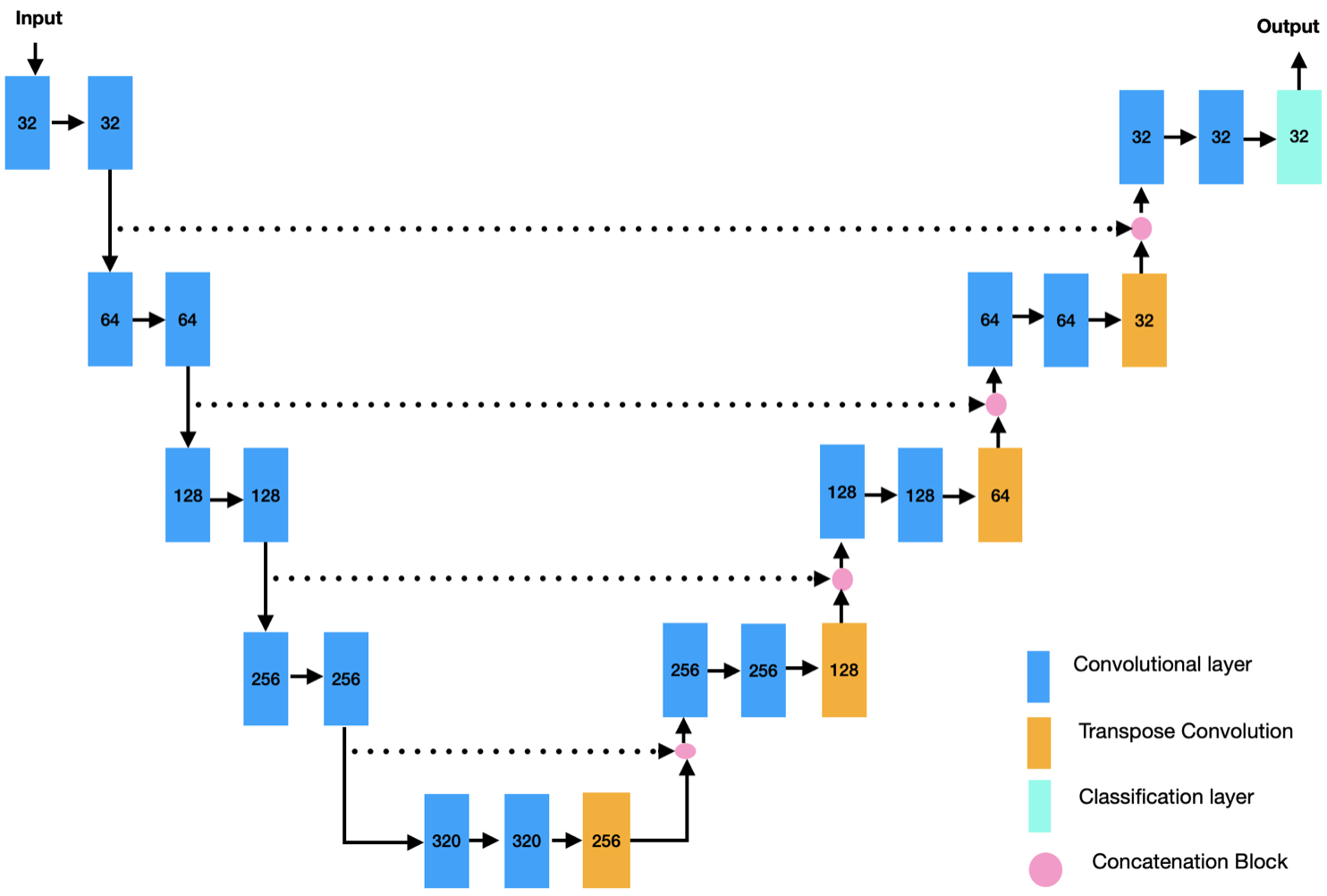}}
\caption{An illustration of the standard Unet architecture use in nnUnet}
\label{fig:Unet}
\end{figure}

The Unet \cite{ronneberger:springer2015} is an end to end architecture for medical image segmentation. It consists of 3 main parts : the encoder, the decoder and bottleneck between the encoder and decoder. The encoder captures contextual information (or features extraction)   and reduces the size of the feature map by half after every convolutional block as we move down the encoding path by implying strided convolutions. 
Pixels localisation is done at the decoder. As we move up the decoder path the size of the feature map is doubled after every convolutional block by implying transposed convolutions, and for the reconstruction process features maps are concatenated to the corresponding map in the encoder path using up-sampling operations.
The bottleneck serves as a bridge, linking the encoding and decoding paths together. It consists of a convolutional block that ensures a smooth transition from the encoder path to the decoder path.
At the encoding path, decoding path and bridge layer each convolutional block consists of a convolutional layer that converts the pixels of the receptive field into a single value before passing it to the next operation followed by an instance normalisation to prevent over-fitting during training, and finally a LeakyReLU activation function to diminish vanishing gradient. A high level diagram to illustrate the architectural structure of the standard Unet is shown in fig \ref{fig:Unet} above.

\subsection{nnUNet Overview}
\label{subsection:nnUNet}
The nnUNet \cite{isensee:NPG2021} is a self-configuring and automatic pipeline for medical image segmentation with the ability to automatically determine and choose the best model hyper-parameters given the data and the hardware availability, thus alleviating the problem of try an error of manual parameters setting.
Given a training data the framework extracts the "data-fingerprint" such as modality, shape, and spacing and base on the hardware (GPU memory) constraints the network topology, image re-sampling methods, and input-image patch sizes are determined. After training is complete the framework determines if post-processing is needed.
During training some parameters are fixed which are : learning rate that is set to 0.01, a maximum training epoch of 1000, the loss function is Cross Entropy plus Dice loss with ADAM as an optimizer, and also data augmentation is done on the fly during training. The framework uses the standard Unet \cite{ronneberger:springer2015} as the network's architecture. 
Please refer to the original publication \cite{isensee:NPG2021} for more information.

\subsection{Residual Connections}
\label{subsection:residual_conne}
Residual connection is a technique use to combat the problem of vanishing gradient developed in \cite{he:IEEECCVPR2016}. The Unet architecture uses the chain rule for back propagation during training. This process can sometimes lead to vanishing gradient and one of the ways to circumvent this, is to introduce residual connection into the network's architecture. The diagram of residual connection is demonstrated on fig \ref{fig:residual_connection} below.

\begin{figure}[h]
\centerline{\includegraphics[width=4cm]{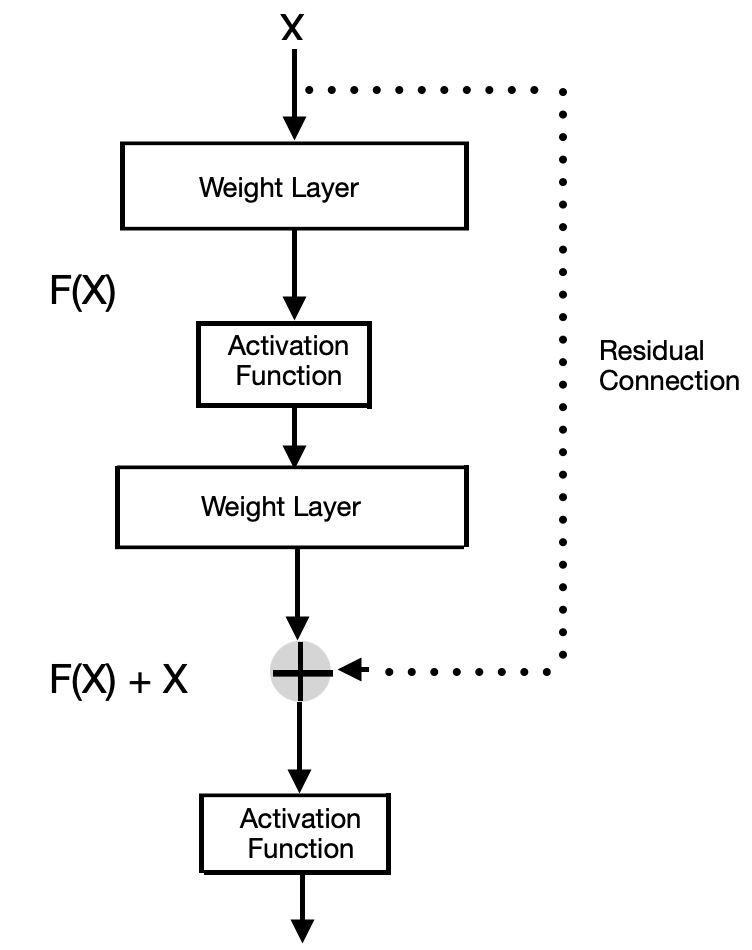}}
\caption{An illustration of the residual connection block to combat the vanishing gradient problem where X is an input and F(X) is a function of X.}
\label{fig:residual_connection}
\end{figure}

\subsection{Atrous Spatial Pyramid Pooling, ASPP}
\label{subsection:aspp}
ASPP is a technique use to extract or capture global contextual features by applying paralleling filters with different frequencies or dilating rate for a given input filter. The diagram of residual connection is demonstrated on fig \ref{fig:ASPP_Structure} below.

\begin{figure}[H]
\centerline{\includegraphics[width=10cm]{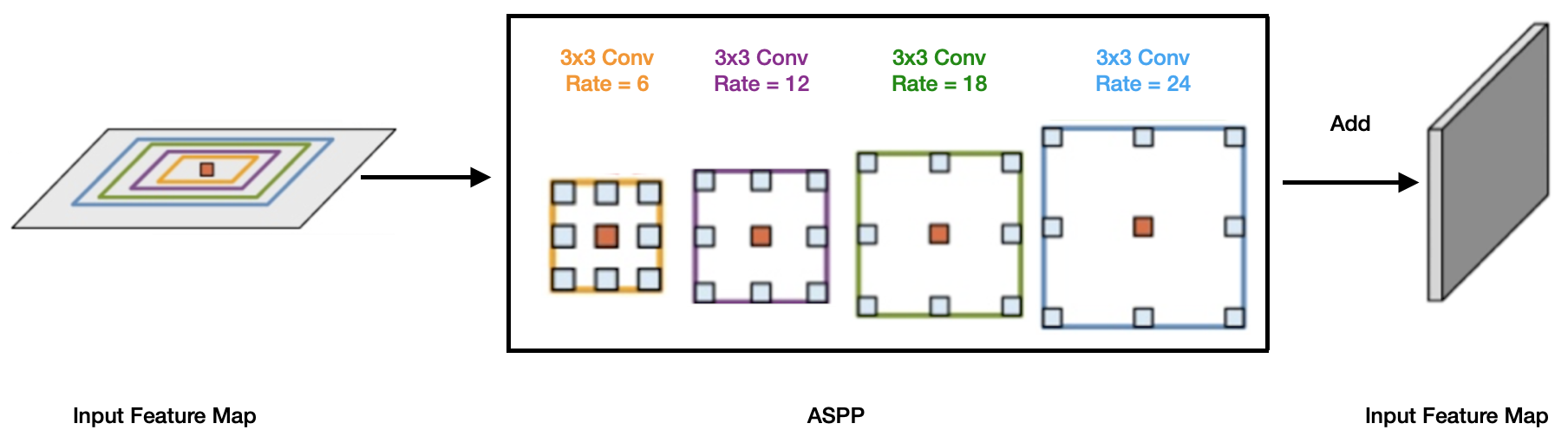}}
\caption{An illustration of multiple parallel filters at different dilating rates or frequencies to capture global information in an ASPP block.}
\label{fig:ASPP_Structure}
\end{figure}

\subsection{nnUnet\_RASPP}
\label{subsection:nnUnet_RASPP}

\begin{figure}[h]
\centerline{\includegraphics[width=11cm]{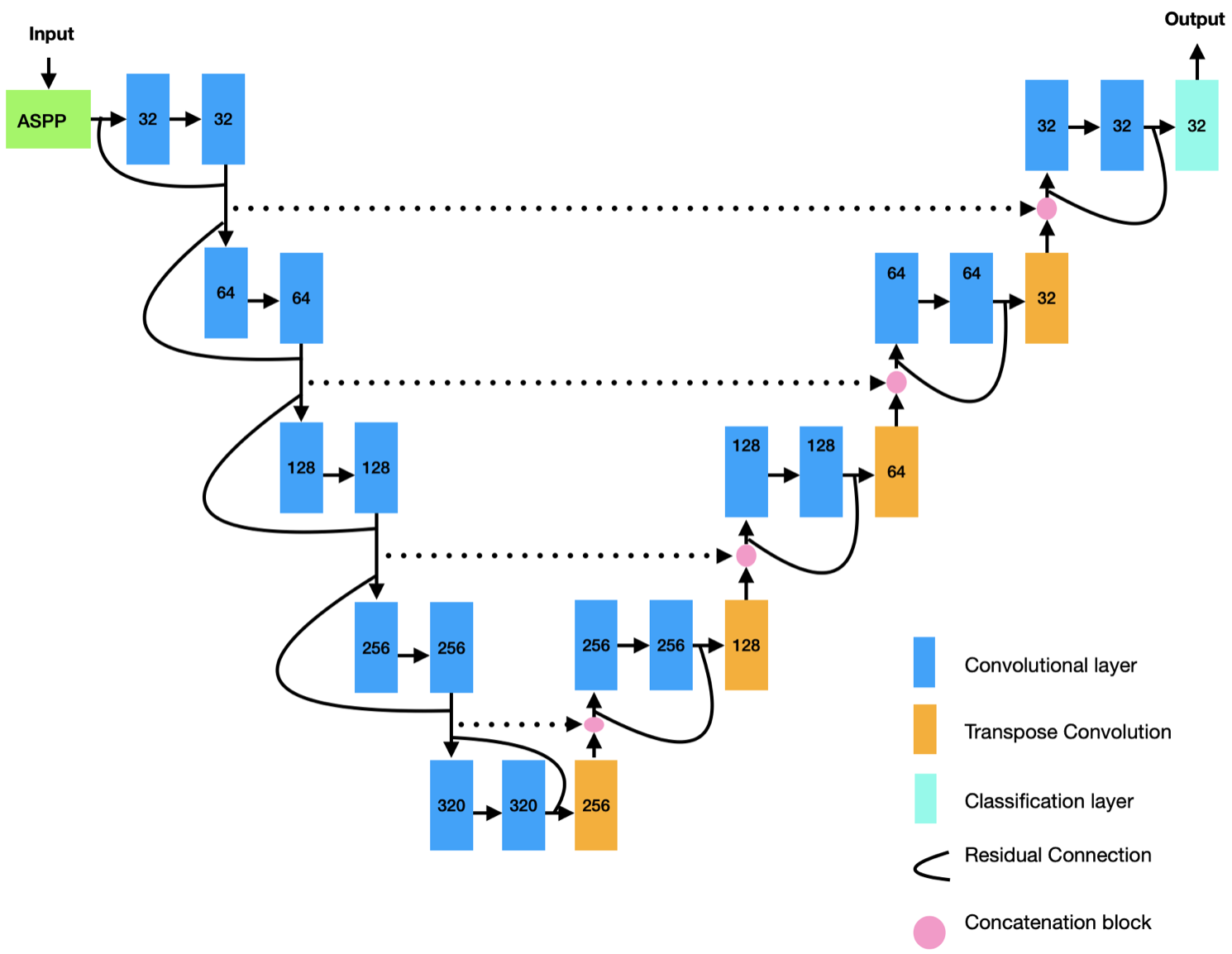}}
\caption{A high level illustration of nnUnet\_RASPP architecture with the ASPP block and the residual connections.}
\label{fig:nnUnet_RASPP}
\end{figure}

Inspired by the success of nnUNet \cite{isensee:NPG2021} we have enhanced the framework by incorporating residual connections and ASPP block in the network's architecture to solve the problem of data source variation. Residual connections were incorporated in every convolutional layer at both the encoding and decoding paths to combat the problem of vanishing gradient and the ASPP was incorporated at the input layer of the encoding path to mitigate the problem of fluid variance.
The diagram of nnUnet\_RASPP is demonstrated in fig \ref{fig:nnUnet_RASPP} above.

\section{Experiments}
\label{section:experiments}
\subsection{Dataset}

\begin{figure}[h]
\centerline{\includegraphics[width=13cm]{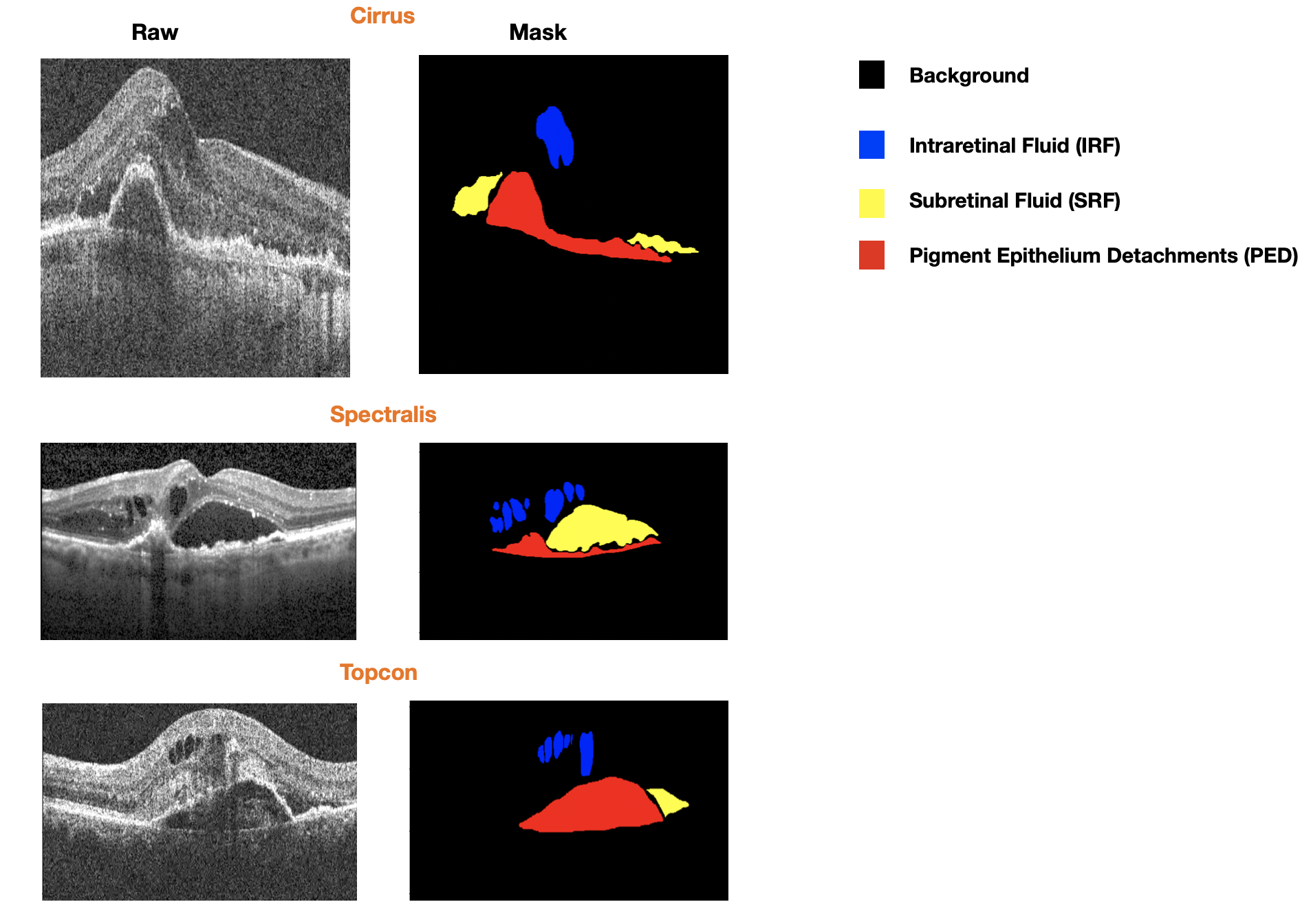}}
\caption{B-Scan examples of raw (column 1) and their corresponded annotated mask (column 2) of OCT volumes taken from the 3 device vendors (rows): Cirrus, Spectralis and Topcon. The classes are coloured as follows : Black for the background,  blue for the Intraretinal Fluid (IRF), yelow for the Subretinal Fluid (SRF) and red for the Pigment Epithelium Detachments (PED). }
\label{fig:dataset_annotation}
\end{figure}

nnUnet\_RASPP was validated on the MICCAI 2017 RETOUCH challenge dataset \cite{bogunovic:IEEE2019}. The dataset is publicly available and it consists of 112 OCT volumes of patients suffering with AMD and DME collected from 3 device  manufacturers: Cirrus, Spectralis and Topcon from 3 clinical centers : Medical University of Vienna (MUV) in Austria, Erasmus University Medical Center (ERASMUS) and Radboud University Medical Center (RUNMC) in The Netherlands.

The dimensions of the OCT volumes per vendor machine are as follows : Each volume of the Cirrus consists 128 B-Scans of $512\times1024$ pixels, Spectralis consists of 49 B-scans of $512\times496$ pixels and 128 B-Scans of $512\times885$ (T-2000) or $512\times650$ (T-1000) pixels for Topcon.

The training set consists of 70 volumes of 24, 24, and 22  acquired with Cirrus, Spectralis, and Topcon, respectively. Both the raw and annotated mask of the training set are made available to the public.
The testing set consists of 42 OCT volumes of 14 volumes per device vendor. The raw or input of the testing set is available publicly but their corresponding annotated masks are held by the organizers of the challenge. Submission and evaluation of prediction on the testing dataset is arranged privately with the organizers and the results are sent to the participants.

Manual annotation was done by 6 grader experts from 2 medical centers : MUV (4 graders supervised by an ophthalmology resident), and RUNMC (2 graders supervised by a retinal specialist). The dataset is annotated for 4 classes of 1 background labelled as 0 and 3 fluids which are : Intraretinal Fluid (IRF) labelled as 1, Subretinal Fluid (SRF) labeled as 2 and Pigment Epithelium Detachments (PED) labelled as 3.

The RETOUCH dataset is particularly interesting because of its high level of variability. It was collected using multiple device vendors, the sizes and number of B-Scans varies per device vendor, and it was collected and annotated in multiple clinical centers. 
Also, for fair comparison the annotated testing set is held by the organizers and submission is curbed to a maximum of 3 per participating team.

\subsection{Training and Testing}
Training was done on the 70 OCT volumes of the training set (both raw and mask volumes). The estimated probabilities and predicted segmentation of the testing set (42 raw volumes) were submitted to the challenge organizers for evaluation on the ground truth or masks. 
Training was done for both nnUnet (as a baseline) and nnUnet\_RASPP architectures. The environmental set up was the same for both architectures.

Also, to further demonstrate the robustness and generalisability of nnUnet\_RASPP, the predicted segmentation of the algorithm was evaluated on OCT volumes from two vendor devices and tested on the third. In this case OCT volumes from the third vendor device weren't seen during training. For this  experiment, two sets of weights were generated which are: (1) Training on 46 OCT volumes from both Spectralis (24 OCT volumes) and Topcon (22 OCT volumes) and evaluated on 14 OCT volumes from the Cirrus testing set and (2) training on
48 OCT volumes from both Cirrus (24 OCT volumes) and Spectralis (24 OCT volumes) and evaluated on 14 OCT volumes from the Topcon testing set. Again the same environmental settings were used to conduct all the experiments.

In the detection task the estimated probabilities of presence of each fluid type is plotted using the receiver operating characteristics (ROC) curve and the area under the curve (AUC) which measures the ability of a binary classifier to distinguish between classes is used as the evaluation 
matrice. The AUC gives a score between 0 and 1 with 1 being the perfect score and 0 is the worst.
For the segmentation task, two evaluation matrices are used to measure the performance of the algorithms which are : (1) the Dice Score (DS) which is twice the intersection, divided by the union. It measures the overlapping of the pixels in the range from 0 to 1 with 1 being the perfect score and 0 being the worst.
(2) The Absolute Volume Difference (AVD) which is the absolute difference between the the predicted and the ground truth. The value ranges from 0 to 1 with 0 being the best result and 1 being the worst. 
The equation to calculate the DS is shown on Eqn~(\ref{eqn:dsc}) and that for AVD  in Eqn~(\ref{eqn:avd}) below.

\begin{equation}
\label{eqn:dsc}
DSC = \frac{ 2 |X \cap Y| } {|X|+|Y|}
\end{equation}

\begin{equation}
\label{eqn:avd}
AVD =  {|X|-|Y|}
\end{equation}

The models were trained on a GPU Server with NVIDIA RTX A6000 48GB.

\subsection{Results}
In this section we report the
performance for the detection task measured by the Area Under the Curve (AUC), and the segmentation task measured by the Dice Score (DS) and Absolute Volume Difference (AVD) for the propose: nnUnet and nnUnet\_RASPP, and compare our results to the current state-of-the-arts architectures.

\begin{figure}[H]

\centerline{\includegraphics[width=12cm]{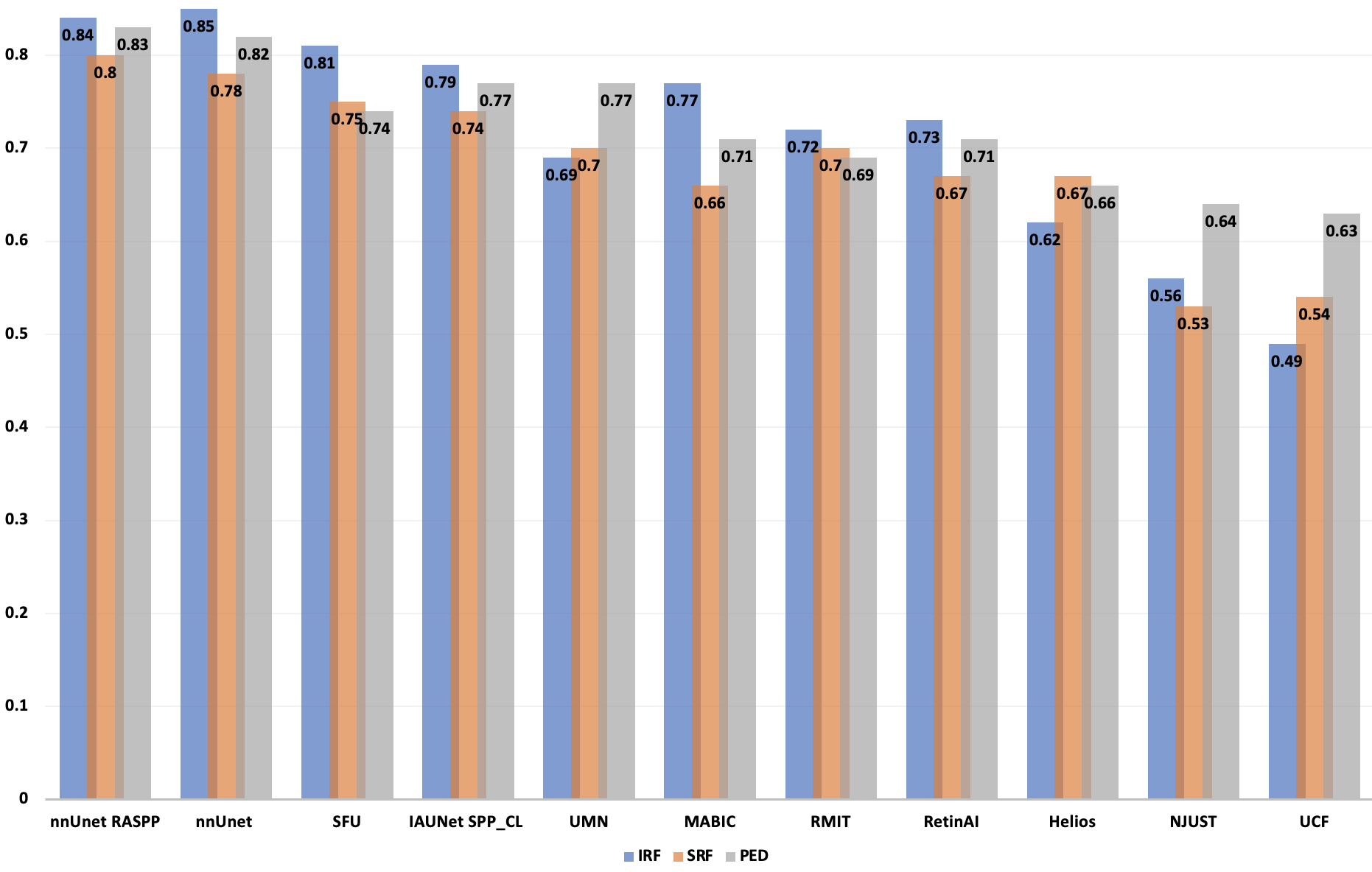}}

\caption{Performance comparison of segmentation measure in DS of the proposed methods: nnUnet\_RASPP and nnUnet, together with the current-state-of-the arts algorithms grouped by the segment classes when trained on the entire 70 OCT volumes of the training set and tested on the holding 42 OCT volumes from the testing set.
}

\label{fig:ds_bar_chart}
\end{figure}

\begin{figure}[H]

\centerline{\includegraphics[width=12cm]{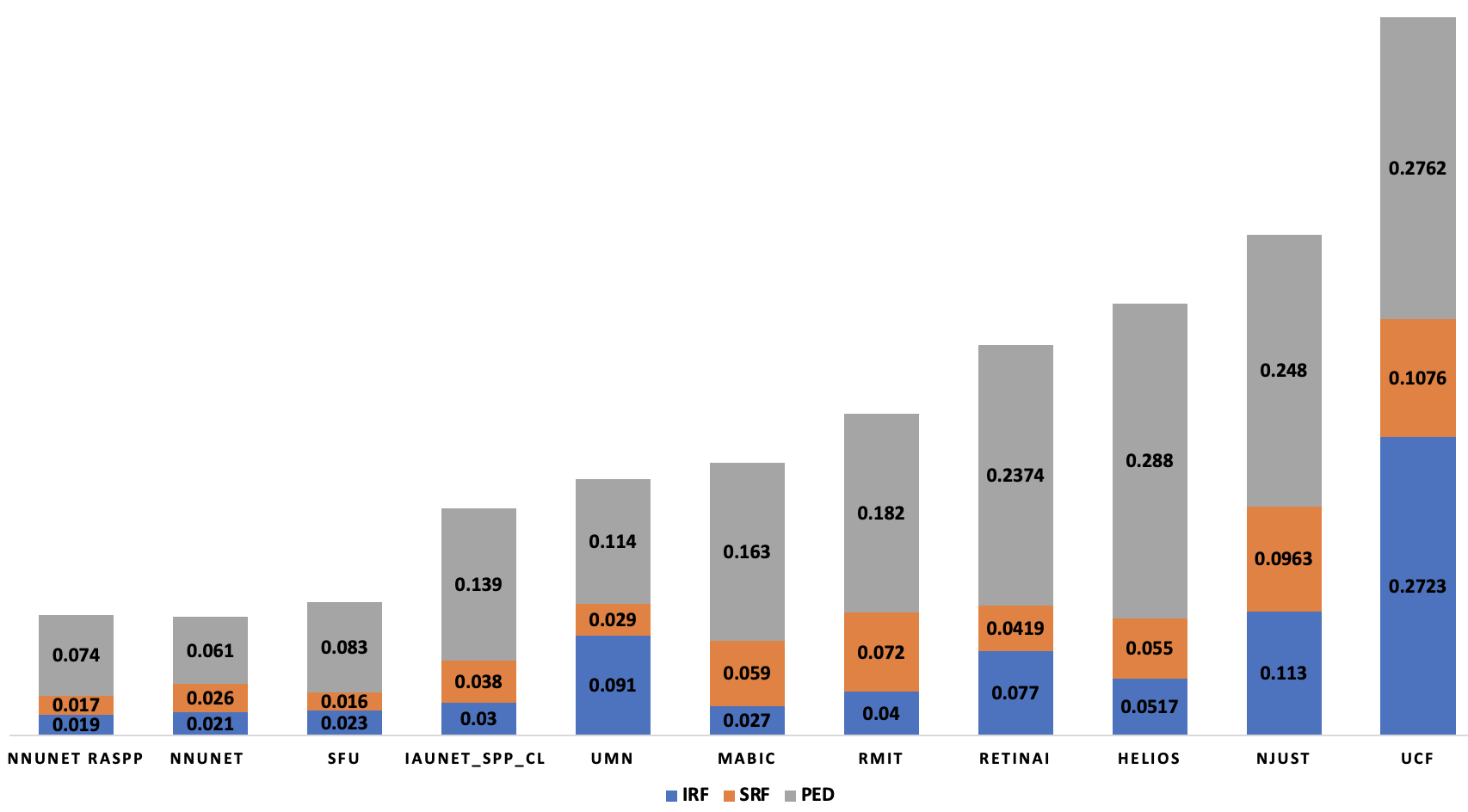}}

\caption{Performance comparison of segmentation measure in AVD of the proposed methods: nnUnet\_RASPP and nnUnet, together with the current-state-of-the arts algorithms grouped by the segment classes when trained on the entire 70 OCT volumes of the training set and tested on the holding 42 OCT volumes from the testing set.
}

\label{fig:avd_bar_chart}
\end{figure}

The segmentation performance grouped by segment classes per algorithm measured in DS is illustrated in Table~\ref{tab:results_ds} with the corresponding diagram in Fig. \ref{fig:ds_bar_chart}, and that measured in AVD is illustrated in Table~\ref{tab:results_avd} with corresponding diagram in Fig. \ref{fig:avd_bar_chart}. 

\begin{figure}[H]

\centerline{\includegraphics[width=12cm]{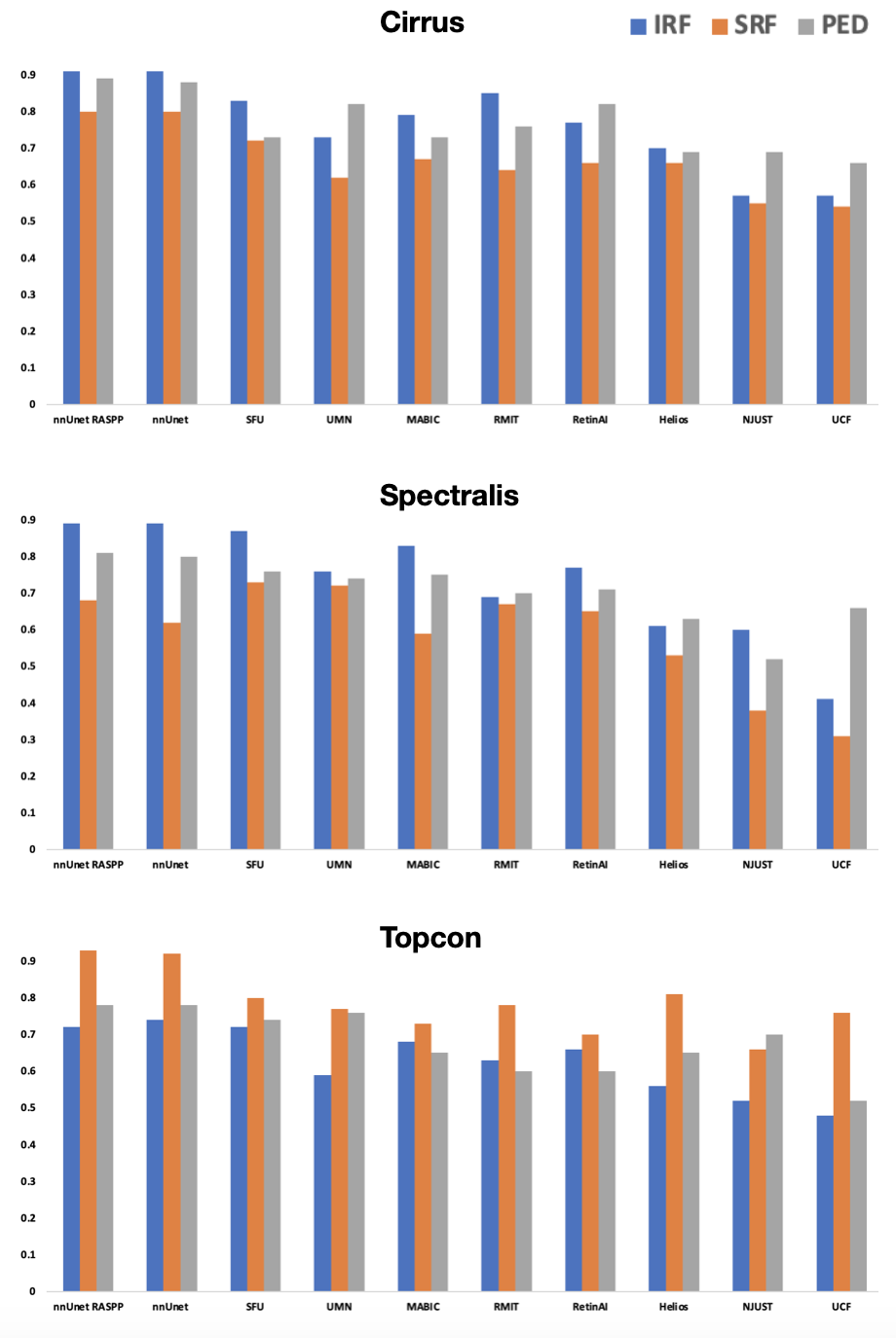}}

\caption{Performance comparison of segmentation measure in DS of the proposed methods: nnUnet\_RASPP and nnUnet, together with the current-state-of-the arts algorithms grouped by the segment classes when trained on the entire 70 OCT volumes of the training set and tested on the holding 42 OCT volumes from the testing set per device.
}

\label{fig:chart_ds_per_device}
\end{figure}

\begin{figure}[H]

\centerline{\includegraphics[width=12cm]{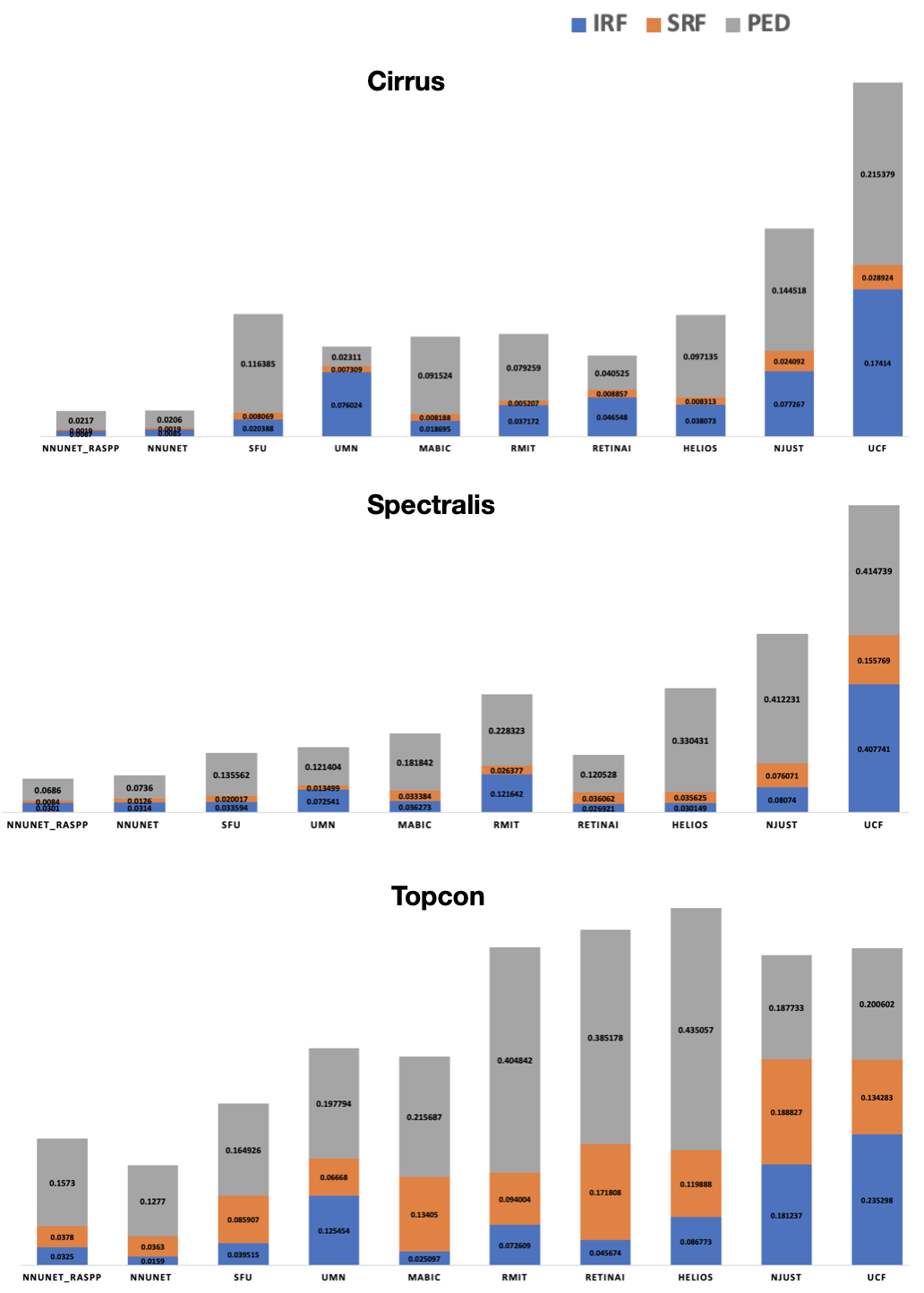}}

\caption{Performance comparison of segmentation measure in AVD of the proposed methods: nnUnet\_RASPP and nnUnet, together with the current-state-of-the arts algorithms grouped by the segment classes when trained on the entire 70 OCT volumes of the training set and tested on the holding 42 OCT volumes from the testing set per device.
}

\label{fig:chart_avd_per_device}
\end{figure}

A detail break down of the DS and AVD per vendor device is shown in 
Table~\ref{tab:results_cirrus_ds_avd_per_device} with the corresponding diagrams of the DS and AVD in Fig. \ref{fig:chart_ds_per_device} and Fig. \ref{fig:chart_avd_per_device} respectively.

Table~\ref{tab:results_dc_avd_cirrus_topcon} with its corresponding diagrams in Fig. \ref{fig:cirrus_topcon_ds} show the results when trained on 2 vendor devices from the training set and tested on the third device from the holding testing set measured in DS. In this case because of the constraint of the evaluation submission (curb to 3 maximum per team) of the predicted segmentation on the testing set, results for nnUnet are unavailable.

\begin{figure}[H]

\centerline{\includegraphics[width=12cm]{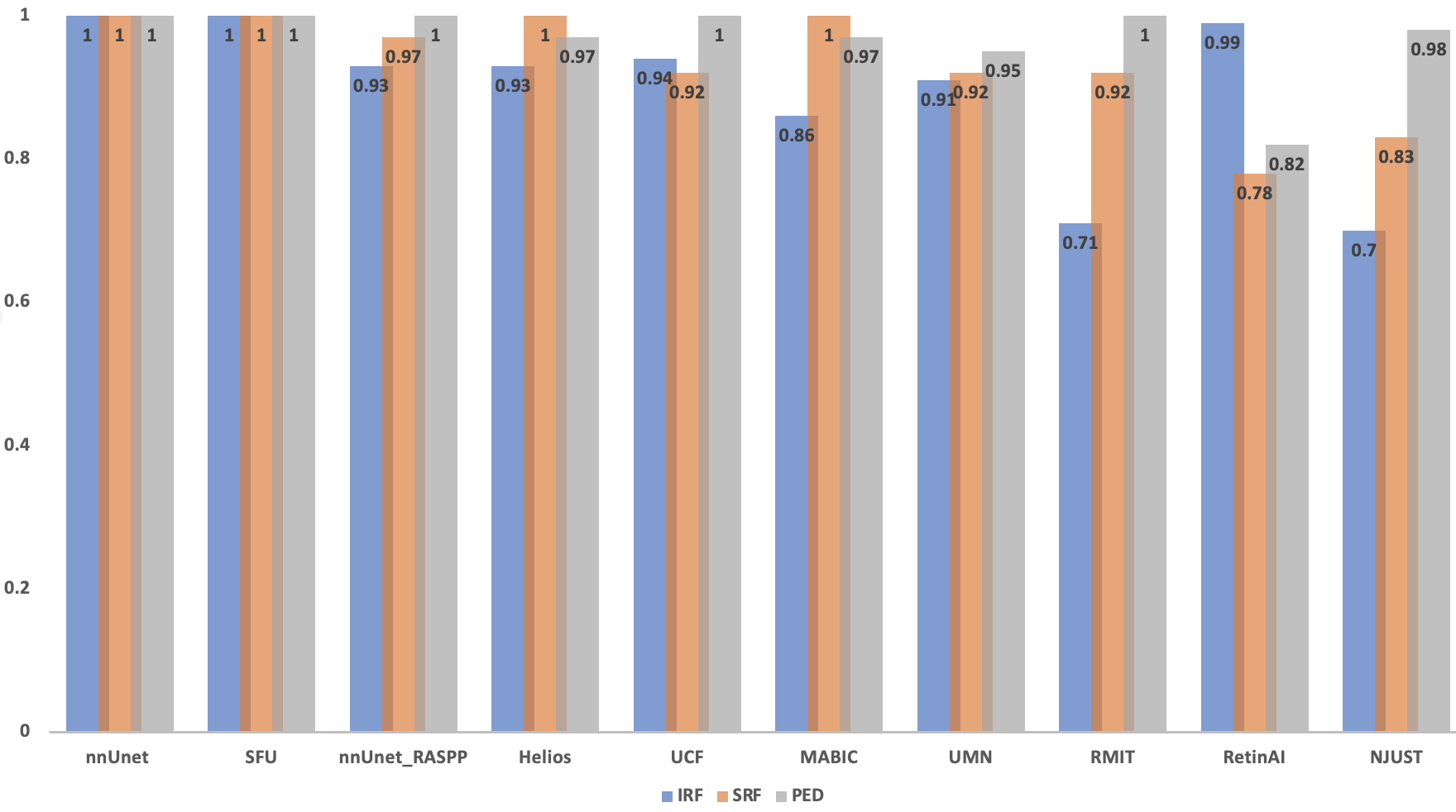}}

\caption{Performance comparison of the detection measure in AUC of the proposed methods: nnUnet\_RASPP and nnUnet, together with the current-state-of-the arts algorithms grouped by the segment classes when trained on the entire 70 OCT volumes of the training set and tested on the holding 42 OCT volumes from the testing set.
}

\label{fig:auc_bar_chart}
\end{figure}

The detection performance grouped by segment classes per algorithm measured by the AUC is illustrated in Table~\ref{tab:results_auc} with the corresponding diagram in Fig. \ref{fig:auc_bar_chart}. 

\begin{figure}[H]
\centering
\centerline{\includegraphics[width=12cm ]{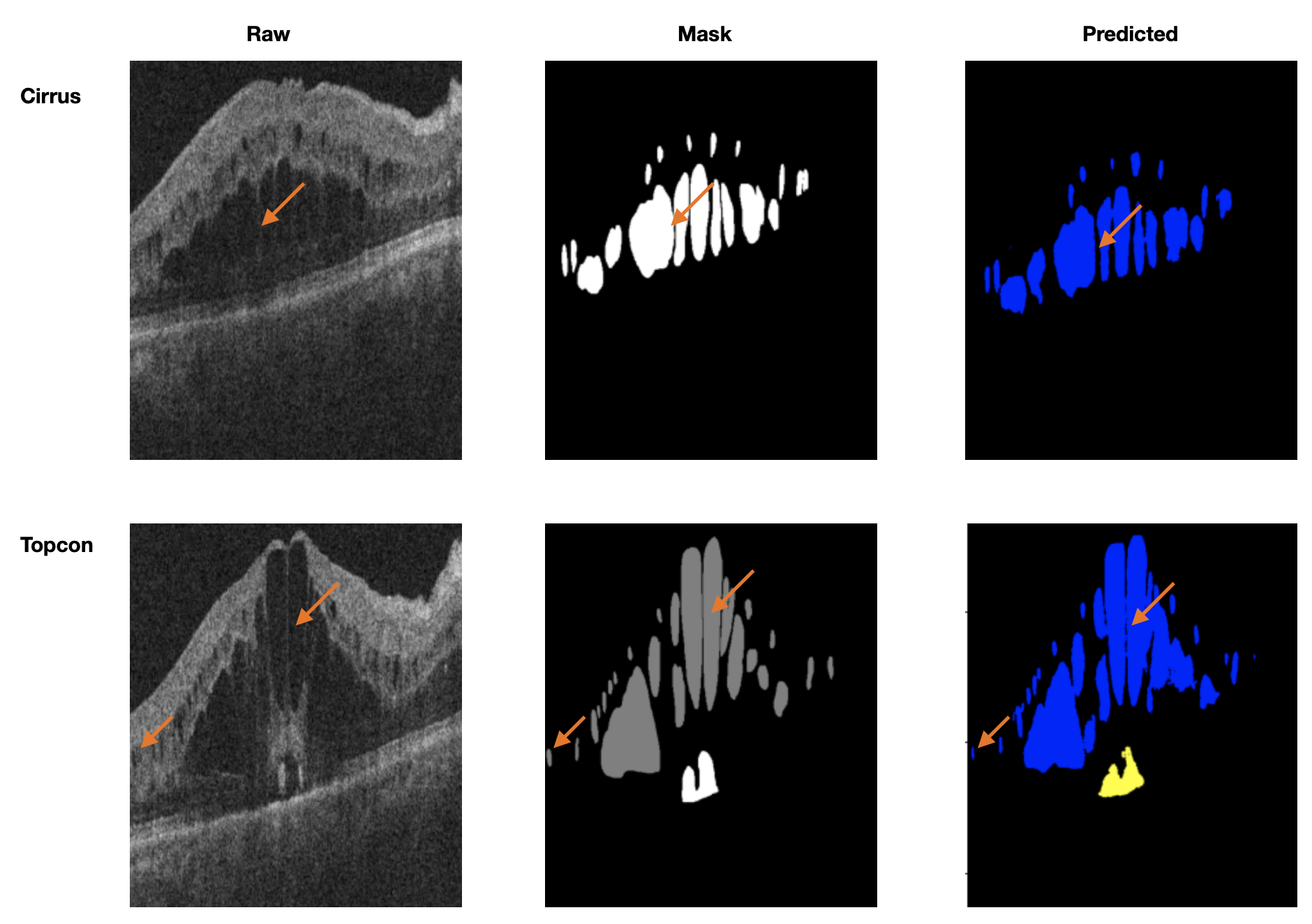}}

\caption{Examples of B-Scans to illustrate the visualization output/predicted of nnUnet\_RASPP, in order of the raw/inputs, mask/annotations and predicted/outputs in columns when trained on the training set of two vendor devices and tested on the training set of the third vendor device (Cirrus and Topcon in row 1 and row 2 respectively). Fine details capture by the model are indicated with orange arrows. }

\label{fig:result_visualization}
\end{figure}

\begin{figure}[H]
\centering
\centerline{\includegraphics[width=12cm ]{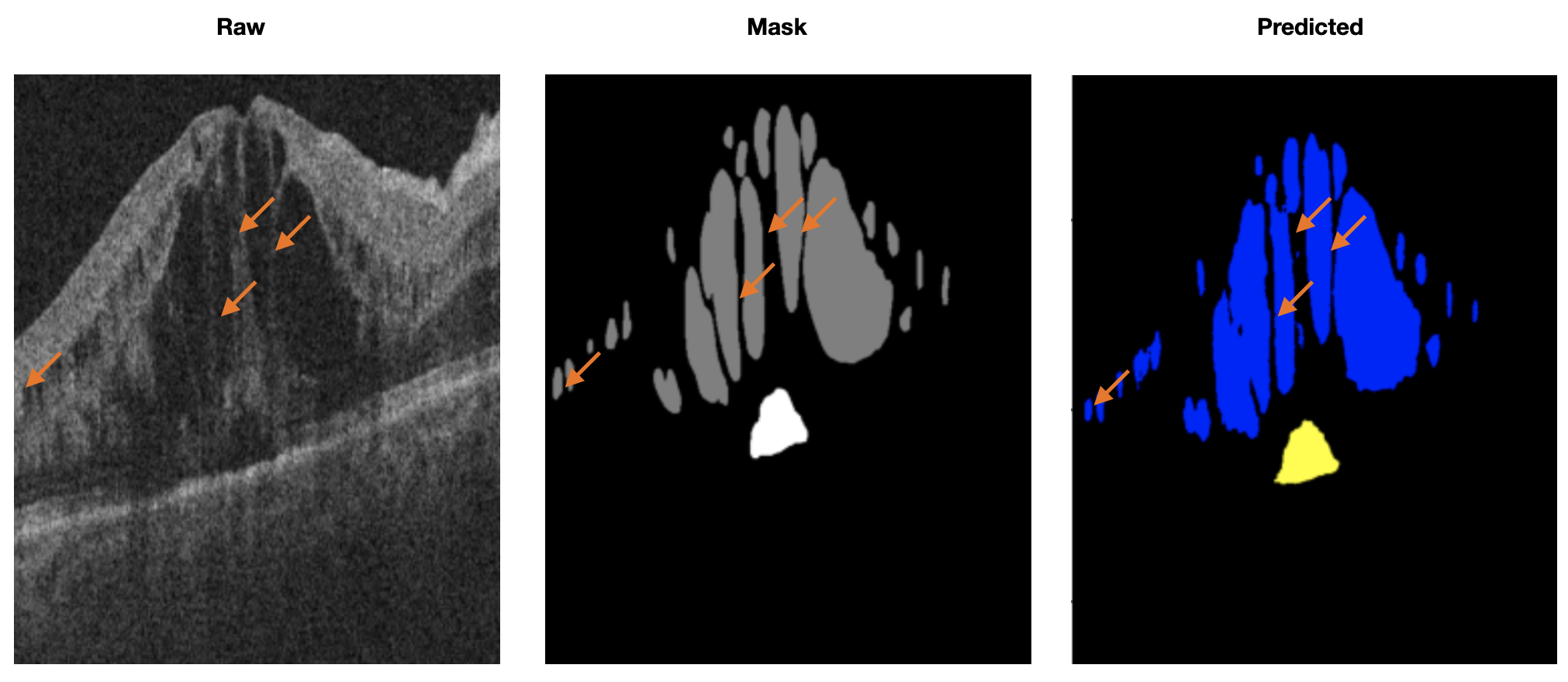}}

\caption{An example of a B-Scan to illustrate the visualization output/predicted of nnUnet\_RASPP, in order of the raw/inputs, mask/annotations and predicted/outputs when zoom out to highlights the fine details capture by the model using orange arrows. This is capture when trained on the training set of the Spectralis and Topcon devices and tested on the training set of the Cirrus device.}

\label{fig:zoom}
\end{figure}

The visualizations using orange arrows to highlight the fine details capture by nnUnet\_RASPP when trained on two vendor devices from the training set and tested on the third from the training set are illustrated in Fig. \ref{fig:result_visualization} and Fig. \ref{fig:zoom}.
From these results, we notice the following: 
\begin{enumerate}
    \item Our propose algorithms: nnUnet\_RASPP and nnUnet outperform the current state-of-the-arts architectures by a clear margin with a mean DS of 0.823 and 0.817 respectively and a mean AVD of 0.036 for nnUnet and 0.041 for nnUnet\_RASPP.
    \item Our propose algorithms: nnUnet obtained a perfect AUC score of 1 for all three fluid classes and 
    nnUnet\_RASPP obtained an AUC score of 0.93, 0.97, and 1.0 for the IRF, SRF, and PED respectively.
    \item Also a detail break down of the DS AND AVD shows both nnUnet\_RASPP and nnUnet clearly outperform the current-state-of-the-arts architectures by a clear margin for all 3 data sources.
    \item We noticed an increase in performance when trained on the training set from two vendor devices and testing on the third vendor testing set scoring a mean DS of 0.84.
    \item Both nnUnet\_RASPP and nnUnet  demonstrate a high level of robustness and generalisability with a constant high level of performance measure in DS and AVD and also when tested on dataset from a third device vendor that is not seen at training. 
    \item nnUnet\_RASPP and nnUnet were the only two algorithms to maintain constant high level performance and generalisability across all 3 data sources.
    \item Dataset acquire from Topcon was the most difficult to segment with nnUnet\_RASPP and nnUnet scoring a mean DS of 0.81 each.
    \item Apart from the IRF class, the nnUnet\_RASPP has the best DS in every single class when compare to the other models/teams.
     \item Further evaluations of the generalisability and high peformance of the propose methods for training on dataset from 2 data sources and tested on the testing set of the third source that isn't seen at training show the nnUnet\_RASPP architecture still outperforms the current state-of-the-arts architectures by a clear margin scoring a mean DS of 0.86 and 0.81 for Cirrus (train on Topcon and Spectralis) and Topcon (train on Cirrus and Spectralis) respectively. In this case, because of the curb of the number of submissions of the predicted segmentation for evaluation to 3 per team we are unable to show the performance of the nnUnet. 

\end{enumerate}

\begin{table}[H]
\addtolength{\tabcolsep}{12pt}

\centering
\begin{tabular}{l c c c c c}
\toprule\toprule
Teams & IRF  &  SRF  & PED  & \thead{Mean}  \\
  \midrule
 
nnUnet\_RASPP      &0.84   & \textbf{0.80}    & \textbf{0.83}    & \textbf{0.823}     \\ 

nnUnet              & \textbf{0.85}  & 0.78   & 0.82    & 0.817     \\

SFU                 & 0.81      & 0.75      & 0.74       & 0.78     \\

IAUNet\_SPP\_CL \cite{xing:IEEE2022}    & 0.79     & 0.74       & 0.77      & 0.77    \\

UMN                 & 0.69     & 0.70       & 0.77     & 0.72      \\

MABIC               & 0.77     & 0.66       & 0.71     & 0.71      \\

RMIT                & 0.72     & 0.70       & 0.69     & 0.70      \\

RetinAI             & 0.73     & 0.67       & 0.71     & 0.70      \\   

Helios             & 0.62     & 0.67       & 0.66     & 0.65      \\ 

NJUST             & 0.56     & 0.53       & 0.64    & 0.58        \\ 

UCF               & 0.49    & 0.54        & 0.63    & 0.55        \\      

\hfill \break
\end{tabular}
\caption{Table of the Dice Scores (DS) by segment classes (columns) and teams (rows) for training on the entire 70 OCT volumes of the training set and tested on the holding 42 OCT volumes from the testing set. }
\label{tab:results_ds}
\end{table}

\begin{table}[h]
\addtolength{\tabcolsep}{12pt}

\centering
\begin{tabular}{l c c c c c}
\toprule\toprule
Teams & IRF  &  SRF  & PED  & \thead{Mean}   \\
  \midrule

nnUnet              & \textbf{0.019}  & 0.017   & 0.074    & \textbf{0.036}    \\

IAUNet\_SPP\_CL \cite{xing:IEEE2022}    & 0.021    & 0.026   & \textbf{0.061}    & \textbf{0.036}   \\

nnUnet\_RASPP      &0.023   & \textbf{0.016}    & 0.083    & 0.041     \\ 

SFU                 & 0.030      & 0.038      & 0.139       & 0.069     \\

UMN                 & 0.091     & 0.029      & 0.114     & 0.078        \\

MABIC               & 0.027     & 0.059       & 0.163    & 0.083        \\

RMIT                & 0.040     & 0.072       & 0.1820     & 0.098    \\

RetinAI             & 0.077     & 0.0419       & 0.2374     & 0.118   \\   

Helios             & 0.0517     & 0.055      & 0.288     & 0.132      \\ 

NJUST             & 0.1130     & 0.0963       & 0.248    & 0.153      \\ 

UCF               & 0.2723    & 0.1076        & 0.2762    & 0.219      \\      
\hfill \break
\end{tabular}
\caption{Table of the Absolute Volume Difference (AVD) by segment classes (columns) and teams (rows) for training on the entire 70 OCT volumes of the training set and tested on the holding 42 OCT volumes from the testing set. }
\label{tab:results_avd}
\end{table}

\begin{table}
\addtolength{\tabcolsep}{10pt}
\begin{center}
\begin{tabular}{ c|cc|cc|cc cc }
\hline
 \multicolumn{7}{c}{\thead{Cirrus}} \\
 \hline
    \hline
Teams    & \multicolumn{2}{c|}{IRF}
            & \multicolumn{2}{c|}{SRF}
                    & \multicolumn{2}{c}{PED}
               \\
    \hline
nnUnet\_RASPP    &   \textbf{0.91}   &   \textbf{0.00670}    &  \textbf{0.80}   &   \textbf{0.00190}   &    \textbf{0.89 }   &   0.021700   \\
\hline

nnUnet     &   \textbf{0.91}   &   0.00850   &  \textbf{0.80}  &  \textbf{0.00190}   &   0.88   &  \textbf{0.02060 }    \\
\hline

SFU   &   0.83  &   0.020388  &   0.72  &   0.008069  &   0.73  &   0.116385   \\
\hline
 
UMN   &   0.73  &   0.076024  &   0.62  &   0.007309  &   0.82  &   0.023110   \\
    \hline

MABIC   &   0.79  &   0.018695  &   0.67  &   0.008188  &   0.73  &   0.091524   \\
    \hline

RMIT   &   0.85  &   0.037172  &   0.64  &   0.005207  &   0.76  &   0.079259   \\
    \hline

RetinAI   &   0.77  &   0.046548  &   0.66  &   0.008857  &   0.82  &   0.040525   \\
    \hline

Helios   &   0.70  &   0.038073  &   0.66  &   0.008313  &   0.69  &   0.097135   \\
    \hline

NJUST   &   0.57  &   0.077267  &   0.55  &   0.024092  &   0.69  &   0.144518   \\
    \hline
UCF   &   0.57  &   0.174140  &   0.54  &   0.028924  &   0.66  &   0.215379   \\
    \hline

\hline
 \multicolumn{7}{c}{\thead{Spectralis}} \\
 \hline
    \hline
Teams    & \multicolumn{2}{c|}{IRF}
            & \multicolumn{2}{c|}{SRF}
                    & \multicolumn{2}{c}{PED}
               \\
    \hline
nnUnet\_RASPP    &   \textbf{0.89}   &   \textbf{0.030100}    &  0.68   &   \textbf{0.008400}   &    \textbf{0.81 }   &   \textbf{0.068600}   \\
\hline

nnUnet     &   \textbf{0.89}   &   0.031400   & 0.62  &  0.012600   &   0.80   &  0.073600    \\
\hline

SFU   &   0.87  &   0.033594  &   \textbf{0.73}  &   0.020017  &   0.76  &   0.135562   \\
\hline
 
UMN   &   0.76  &   0.072541  &   0.72  &   0.013499  &   0.74  &   0.121404   \\
\hline

MABIC   &   0.83  &   0.036273  &   0.59  &   0.033384  &   0.75  &   0.181842   \\
\hline

RMIT   &   0.69  &   0.121642  &   0.67  &   0.026377  &   0.70  &   0.228323   \\
\hline

RetinAI   &   0.77  &   0.026921 &   0.65  &   0.036062  &   0.71  &   0.120528   \\
\hline

Helios   &   0.61  &   0.030149  &   0.53  &   0.035625  &   0.63  &   0.330431   \\
\hline

NJUST   &   0.60  &   0.080740  &   0.38  &   0.076071  &   0.52  &   0.412231   \\
    \hline
UCF   &   0.41  &   0.407741  &   0.31  &   0.155769  &   0.52  &   0.414739   \\
    \hline

\hline
 \multicolumn{7}{c}{\thead{Topcon}} \\
 \hline
    \hline
Teams    & \multicolumn{2}{c|}{IRF}
            & \multicolumn{2}{c|}{SRF}
                    & \multicolumn{2}{c}{PED}
               \\
    \hline
nnUnet\_RASPP    &   0.72   & 0.032500    &  \textbf{0.93}   &  0.037800   &    \textbf{0.78 }   &   0.157300    \\
\hline

nnUnet     &   \textbf{0.74}   &    \textbf{0.015900}   & 0.92  &  \textbf{0.036300}   &   \textbf{0.78 }   &  \textbf{0.127700}    \\
\hline

SFU   &   0.72  &   0.039515  &   0.80  &   0.085907  &   0.74  &   0.164926   \\
\hline
 
UMN   &   0.59  &   0.125454  &   0.77  &   0.066680  &   0.76  &   0.197794   \\
\hline

MABIC   &   0.68  &   0.025097  &   0.73  &   0.134050  &   0.65  &   0.215687   \\
\hline

RMIT   &   0.63  &   0.072609  &   0.78  &   0.094004  &   0.60  &   0.404842   \\
\hline

RetinAI   &   0.66  &   0.045674 &   0.70  &   0.171808  &   0.60  &   0.385178   \\
\hline

Helios   &   0.56  &   0.086773  &   0.81  &   0.119888  &   0.65  &   0.435057   \\
\hline

NJUST   &   0.52  &   0.181237  &   0.66  &   0.188827  &   0.70  &   0.187733   \\
    \hline
UCF   &   0.48  &   0.235298  &   0.76  &   0.134283  &   0.61  &   0.200602   \\
    \hline

\end{tabular}
    \end{center}
\caption{Table of the Dice Score (DS) and Absolute Volume Difference (AVD) by segment classes (columns) and teams (rows) for training on the entire 70 OCT volumes of the training set and tested on the holding 42 OCT volumes from the testing set per device. }
\label{tab:results_cirrus_ds_avd_per_device}
\end{table}

\begin{table}
\addtolength{\tabcolsep}{2pt}
\begin{center}
\begin{tabular}{ c|cc|cc|cc|cc cc }
\hline
 \multicolumn{9}{c}{\thead{Cirrus}} \\
 \hline
    \hline
Teams    & \multicolumn{2}{c|}{IRF}
            & \multicolumn{2}{c|}{SRF}
                    & \multicolumn{2}{c}{PED}
                     & \multicolumn{2}{c}{Mean}
               \\
    \hline
nnUnet\_RASPP    &   \textbf{0.90}   &   \textbf{0.0122}    &  \textbf{0.78}   &   \textbf{0.0031}   &    \textbf{0.89 }   &   \textbf{0.019}    & \textbf{0.86 }   &   \textbf{0.0114}\\
\hline

SFU   &   0.83  &   0.0204  &   0.72  &   0.0081  &   0.73  &   0.1164   &  0.76   &   0.0483 \\
\hline
 
UMN   &   0.73  &   0.0760  &   0.62  &   0.0073  &   0.82  &   0.0231   & 0.72   &   0.0355 \\
    \hline

MABIC   &   0.79  &   0.0187  &   0.67  &   0.0082  &   0.73  &   0.0915  & 0.73   &   0.0395  \\
    \hline

RMIT   &   0.85  &   0.0372  &   0.64  &   0.0052  &   0.76  &   0.0793   & 0.75   &   0.0406 \\
    \hline

RetinAI   &   0.77  &   0.0466  &   0.66  &   0.0089  &   0.82  &   0.0405  & 0.75   &   0.0320  \\
    \hline

Helios   &   0.70  &   0.0381  &   0.66  &   0.0083  &   0.69  &   0.0971  &  0.68   &   0.0478 \\
    \hline
    
SVDNA \cite{koch:MICCAI2022}   & 0.61    & --   & 0.66   & --    & 0.74     & --          &   0.67  & -- \\
     \hline
NJUST   &   0.57  &   0.0773  &   0.55  &   0.0241  &   0.69  &   0.1446   & 0.60   &   0.0820 \\
    \hline
UCF   &   0.57  &   0.1741  &   0.54  &   0.0289  &   0.66  &   0.2154   & 0.59   &   0.1395 \\
    \hline

\hline
 \multicolumn{9}{c}{\thead{Topcon}} \\
 \hline
    \hline
Teams    & \multicolumn{2}{c|}{IRF}
            & \multicolumn{2}{c|}{SRF}
                    & \multicolumn{2}{c}{PED}
                    & \multicolumn{2}{c}{Mean}
               \\

    \hline
nnUnet\_RASPP    &   0.72   & \textbf{0.0201}     &  \textbf{0.93}   &  \textbf{0.0298}   &    \textbf{0.78 }   &   \textbf{0.2119}   &    \textbf{0.81 }   &   \textbf{0.0873}   \\
\hline

SFU   &   0.72  &   0.0395  &   0.80  &   0.0859  &   0.74  &   0.1649  &   0.75  &  0.0968\\
\hline
 
UMN   &   0.59  &   0.1255  &   0.77  &   0.0667  &   0.76  &   0.1978   &   0.71  &  0.1300 \\
\hline

SVDNA \cite{koch:MICCAI2022}   & 0.61    & --   & 0.80   & --    & 0.72     & --          &   0.71  & -- \\
     \hline
     
MABIC   &   0.68  &   0.0251  &   0.73  &   0.1341  &   0.65  &   0.2157    &  0.69  &  0.1250\\
\hline

RMIT   &   0.63  &   0.0726  &   0.78  &   0.0940  &   0.60  &   0.4048    &   0.67  &  0.1905\\
\hline

RetinAI   &   0.66  &   0.0457 &   0.70  &   0.1718  &   0.60  &   0.3852    &   0.65  &  0.2009\\
\hline

Helios   &   0.56  &   0.0868  &   0.81  &   0.1199  &   0.65  &   0.4351    &   0.67  &  0.2139 \\
\hline

NJUST   &   0.52  &   0.1812  &   0.66  &   0.1888  &   0.70  &   0.1877    &   0.63  &  0.1859\\
    \hline
UCF   &   0.48  &   0.2353  &   0.76  &   0.1343  &   0.61  &   0.2006   &   0.62  &  0.1900 \\
    \hline

\end{tabular}
    \end{center}
\caption{Table of the DS and AVD by segment classes (columns) and teams (rows) trained on 48 OCT volumes from 2 device sources and evaluated on 14 OCT volumes from the testing set on the third device that wasn't seen at training. }
\label{tab:results_dc_avd_cirrus_topcon}
\end{table}

\begin{figure}[H]

\centerline{\includegraphics[width=12cm]{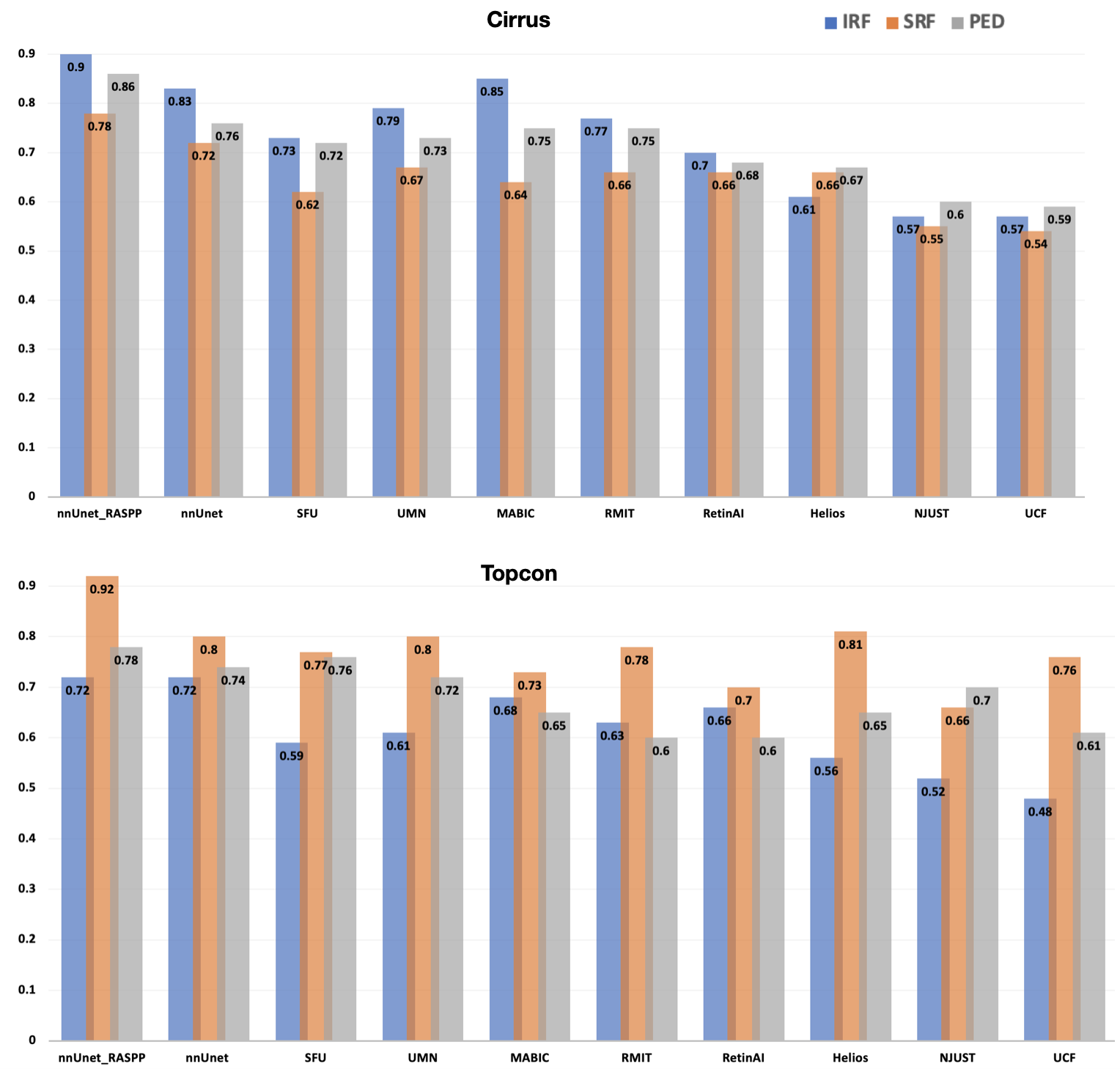}}

\caption{Performance comparison of segmentation measure in DS of the propose nnUnet\_RASPP, together with the current-state-of-the arts algorithms group by the segment classes train on 46 OCT volumes from both Spectralis (24 OCT volumes) and Topcon (22 OCT volumes) and evaluated on the holding testing set (cirrus top and Topcon below).}
\label{fig:cirrus_topcon_ds}
\end{figure}

\begin{figure}[H]

\centerline{\includegraphics[width=12cm]{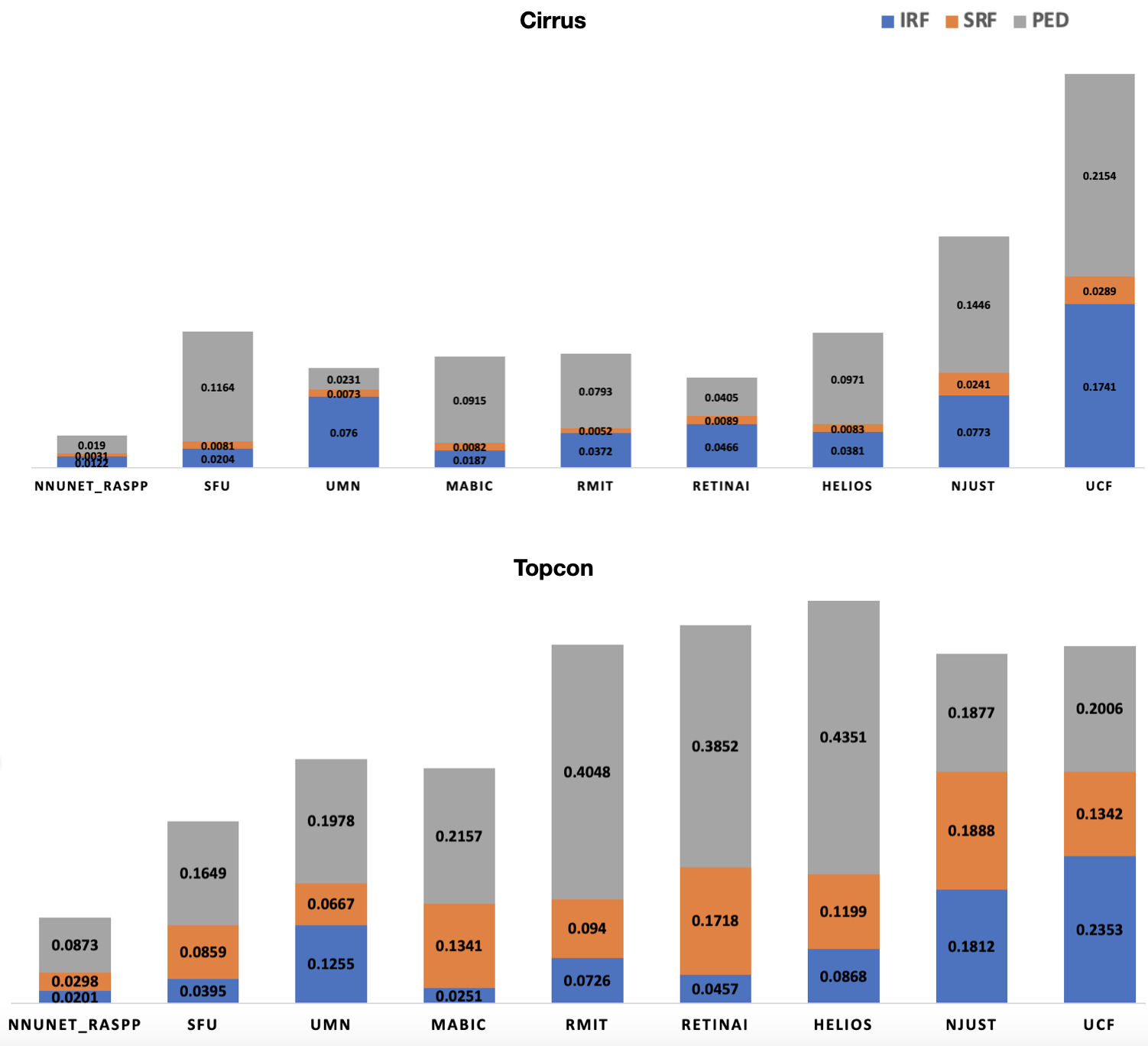}}

\caption{Performance comparison of segmentation measure in AVD of the propose nnUnet\_RASPP, together with the current-state-of-the arts algorithms group by the segment classes train on 46 OCT volumes from both Spectralis (24 OCT volumes) and Topcon (22 OCT volumes) and evaluated on the holding testing set (cirrus top and Topcon bottom).}
\label{fig:cirrus_topcon_ds}
\end{figure}


\begin{table}[H]
\addtolength{\tabcolsep}{12pt}

\centering
\begin{tabular}{l c c c c c}
\toprule\toprule
Teams & IRF  &  SRF  & PED  & \thead{Mean}  \\
  \midrule
 
nnUnet      &\textbf{1.0}  & \textbf{1.0}    & \textbf{1.0}    & \textbf{1.0}     \\ 
SFU         &\textbf{1.0}  & \textbf{1.0}    & \textbf{1.0}    & \textbf{1.0}     \\

nnUnet\_RASPP     & 0.93      & 0.97   & \textbf{1.0}    & 0.97    \\

Helios            & 0.93     & \textbf{1.0}    & 0.97     & 0.97     \\ 

UCF               & 0.94    & 0.92        & \textbf{1.0}    & 0.95    \\  

MABIC              & 0.86     & \textbf{1.0}      & 0.97    & 0.94      \\

UMN                & 0.91     & 0.92       & 0.95     & 0.93      \\

RMIT                & 0.71     & 0.92      & \textbf{1.0}     & 0.88     \\

RetinAI             & 0.99     & 0.78       & 0.82     & 0.86      \\   

NJUST             & 0.70     & 0.83       & 0.98    & 0.84        \\

\hfill \break
\end{tabular}
\caption{Table of the Area Under the Curve (AUC) by segment classes (columns) and teams (rows) for training on the entire 70 OCT volumes of the training set and tested on the holding 42 OCT volumes from the testing set. }
\label{tab:results_auc}
\end{table}

\section{Conclusions}
\label{section:conclusions}

In this work, we have investigated the problem of detection and segmentation of multiple fluids in retinal OCT volumes  acquired from multiple device vendors. Inspired by the success of of the nnUNet \cite{isensee:NPG2021} we have enhanced the model's architecture to build a novel algorithm call nnUnet\_RASPP. Both nnUNet and nnUnet\_RASPP were evaluated on the MICCAI 2017 RETOUCH challenge dataset \cite{bogunovic:IEEE2019}. We submitted predictions for both architectures and experimental results show that for the current league table and other known published results our algorithms outperform the current state-of-the-arts architectures by a clear margin as they occupy the first and second places for the DS, AVD and AUC evaluation.

Our main contribution is : we have enhanced the nnUNet by incorporating the residual blocks and ASPP block into the network's architecture to solve this particular problem.

The propose algorithms provide useful information for further diagnosis and monitoring the progress of retinal diseases such as AMD, DME and Glaucoma. In the future we look to investigate our algorithms on other medical image datasets once they are available publicly.

\clearpage
\bibliography{references}

\end{document}